\documentclass{elsart}
\usepackage{amsmath}
\usepackage{graphicx}
\usepackage{psfrag}

\newcommand{\beq}{\begin{equation}}
\newcommand{\eeq}{\end{equation}}
\newcommand{\beqa}{\begin{eqnarray}}
\newcommand{\eeqa}{\end{eqnarray}}
\newcommand{\Bint}{B_{\text{int}}}
\DeclareMathOperator{\Tr}{Tr}

\begin{document}
\begin{frontmatter}
\title{Decoherence effects on the tunneling rate of paramagnetic and superparamagnetic particles}
\author{E. H. {Martins Ferreira}},
\author{M. C. Nemes},
\author{H.-D. Pfannes},
\address{Departamento de F\'\i sica, Universidade Federal de Minas
Gerais, CP 702, 30123-970 Belo Horizonte, MG, Brazil}

\begin{abstract}
We analyze the effects of the environment on the spin tunneling process of paramagnetic and superparamagnetic particles and conclude that the ``bare'' macroscopic tunneling rate is hardly affected in such case, but others more effective processes come up to change the magnetization state of the particle. We measure the degree of coherence loss by the linear entropy which will account for all correlations present in the process. We conclude that for both, paramagnetic and superparamagnetic particles, the decoherence time scale is extremely short ($\sim 10^{-8..-16}s$), indicating that coherent tunneling should be strongly suppressed in favor of incoherent tunneling, i.e., the population of higher levels with subsequent decay. Interestingly enough, the ground state tunneling rate is hardly affected by such dissipative mechanisms. However other processes immediately give rise to new (incoherent) possibilities of crossing the barrier.
\end{abstract}

\end{frontmatter}

Recently discovered magnetic molecules\cite{kahn,gatt,can,sessoli} are of interest because of the gigantic relaxation times of their magnetization. Recent experiments on magnetic relaxation of molecular crystals of Mn$_{12}$ and Fe$_8$ find strong evidence for tunneling mediated relaxation at low temperatures\cite{paulsen,novak,friedman,thomas,hern,sang}. Several theoretical studies have been performed\cite{enz,hemmen,leuenb1,stamp}, leading to tunneling rates of the order of a few months. The situation with superparamagnetic particles is less clear, however. A change in magnetization direction at low temperatures is indeed observed, the question remaining as to whether this is due to a unitary tunneling effect. Contradictory statements regarding the observability of the tunneling effect can be found in the literature\cite{pfannes1,pfannes2}.

In the present letter, we set up a collective model for the tunneling of magnetic molecules, derive the corresponding tunneling rates and include dissipation via coupling with a Caldeira-Leggett type of environment. As expected, the decoherence time scale is found to be much shorter than the characteristic tunneling time. However, surprisingly enough, we find that the tunneling rate is hardly affected by the environment induced dissipation. We therefore conclude, within the scope of the present model, that the spin-phonon interaction does not significantly interfere with the coherences which are essential for the tunneling process. It however very quickly opens up several alternative routes for the system to change its magnetization, rendering the unitary tunneling mechanism but a possibility among many others. Our result strongly supports the picture of incoherent tunneling, since coherence is destroyed just about immediately. 

We start by considering a Hamiltonian which describes the magnetization properties of such molecules\cite{leuenb2},
\beq
H_S = - AS_z^2 + B(S_+^2 + S_-^2). \label{eq:ham}
\eeq
The anisotropy constants satisfy $A \gg B > 0$. $S_z$ and $S_\pm = S_x \pm iS_y$ are spin operators. The energy splitting of the ground state due to the second term on the r.h.s. of Eq.~(\ref{eq:ham}) can be calculated by means of path integral techniques and the instanton method. Details of this calculation can be found in Ref.~\cite{erlon}. We get
\beq
\Delta E_{\text{inst}} = \frac{8As^{3/2}}{\sqrt{\pi}}\left(\frac{B}{A}\right)^s \cos(\pi s).
\eeq
This result can be compared with that obtained by Hartmann-Boutron by means of others methods\cite{hart}. For magnetic molecules, like Mn$_{12}$, $s=10$ and for typical superparamagnetic particles, $s \sim 3000\hbar$. For both we have $B \ll A$, and therefore $\Delta  E_{\text{inst}} \sim 0$, suggesting then that th observation of this phenomenon in the laboratory is impossible.

Obviously thermally activated over-barrier turnover of the magnetization direction at sufficiently high temperatures (normal superparamagnetism) is possible. In the following, however, we analyze the possibility of spin-tunneling, i.e., under-barrier changes of the magnetization direction.

The model of a single collective variable is apparently too naive. We therefore next take into account the fact that the magnetic particles are not isolated and estimate the characteristic time of environmental effects. We consider phonon degrees of freedom, represented as a set of harmonic oscillators described by the Hamiltonian
\beq
H_R = \sum_\alpha \hbar\omega_\alpha\ a_\alpha^\dagger a_\alpha
\eeq
and coupled to $H_S$ (this model has been proposed in \cite{pfannes3})
\beq
H_{\text{int}} = \Bint(S_+^2 + S_-^2) \sum_{\alpha}
\sqrt{\frac{\hbar}{2M\omega_\alpha}}k_\alpha(a_\alpha^\dagger + a_\alpha),
\eeq
where $\Bint$ is an overall interaction constant, $M$ the particle mass, $k_\alpha$ the wave number associated to the phonon with frequency $\omega_\alpha$ and $a_\alpha^\dagger(a_\alpha)$ creation (annihilation) operators of the environmental degrees of freedom.

Assuming this interaction to be weak enough with respect to the collective dynamics, Eq.~(\ref{eq:ham}), we use perturbation theory to evaluate the reduced density of the spin degree of freedom,
\beqa
\tilde\rho(t) & = &  \tilde\rho(0) - \frac{i}{\hbar}\int_0^t dt'[\tilde
  H_{\text{int}}(t'),\tilde\rho(0)] \nonumber \\
& & - \frac{1}{\hbar^2}\int_0^t dt'\int_0^{t'}dt''[\tilde
  H_{\text{int}}(t'),[\tilde H_{\text{int}}(t''),\tilde\rho(0)]] + \cdots
    \nonumber \\
& = & \tilde\rho^{(0)} + \tilde\rho^{(1)} + \tilde\rho^{(2)} + \cdots
\eeqa
where $\tilde H_{\text{int}} = \e^{iH_0t}H_{\text{int}}\e^{-iH_0t}$ and $H_0 \equiv H_S + H_R$. The reduced density is then 
\beq
\tilde\rho_S = \Tr_R[\tilde\rho].
\eeq
$\Tr_R$ represents the trace over the reservoir (phonons) variables. From the above expressions it is not difficult to obtain the two quantities which lead us to our final conclusions:

\begin{enumerate}

\item The tunneling probability, i.e., the probability as a function of time to
find our system in state $|-s\rangle$, having been initially prepared in
$|+s\rangle$
\beq \begin{split}
{P}_{-s}(t) & = \Tr[|-s\rangle\langle-s|\tilde\rho_S(t)] \\
& = \sin^2\left(\frac{\Delta \omega_{\text{inst}} t}{2}\right) \left[1 - \frac{18\Bint^2s_{-2}} {\hbar M c^2\omega_D^3} \left(\frac{(\omega_{s,s-2})^3}{\e^{\hbar\omega_{s,s-2}/kT}-1} \right)t\right] \label{eq:P}
\end{split}\eeq

\item
The coherence loss as measured by the linear entropy (or idempotency defect
\cite{zurek})
\beq\begin{split}
\delta_S(t) & = 1 - \Tr_S[(\tilde\rho_S)^2] \\
& = \frac{3\Bint^2}{4\pi^2\hbar\rho c^5} \left[\frac{s_{+2}^m(\omega_{m,m+2})^3}{1-\e^{-\hbar\omega_{m,m+2}/k_BT}} +\frac{s_{-2}^m(\omega_{m,m-2})^3}{1-\e^{-\hbar\omega_{m,m-2}/k_BT}}\right]t \label{eq:delta}
\end{split}\eeq
From this formula we also define the ``decoherence time'' ($\tau_{\text{dec}}$) as being the inverse of the factor multiplying $t$, or in others words, the time at which $\delta_S(t)=1$.
\end{enumerate}

In the above expressions, we have used a spectral function for the phonons
given by
\beq
J(\omega) = \frac{\pi}{2}\Bint^2\sum_\alpha
\frac{k_\alpha^2}{M\omega_\alpha}\delta(\omega-\omega_\alpha) \label{eq:J}
\eeq
and a linear dispersion relation for the phonons $\omega_\alpha = ck_\alpha$,
where $c$ is the sound velocity in the medium. Moreover, Debye's model is used
to transform the sum into an integral, i.e.
\[
\sum_\alpha Q(\omega_\alpha) \rightarrow \int_0^{\omega_D} d\omega
  g(\omega)Q(\omega)
\]
with $g(\omega) = 9\omega^2/\omega_D^3$ and $\omega_D^{-3} = 3V/(18\pi^2c^3)$,
$V$ being the particle's volume. Nevertheless it is only necessary to know the value of the density $\rho$ since $V$ and $M$ always appear in formulas in the form $M/V$. With these ingredients, Eq.~(\ref{eq:J})
becomes
\beq
J(\omega) = \frac{9\pi\Bint^2\omega^3}{2Mc^2\omega_D^3}
\eeq
corresponding to superohmic dissipation, according to Caldeira and Leggett's
model\cite{leggett}. In Eq.~(\ref{eq:delta}), $s_{\pm 2}^{m} \equiv (s \mp m)(s 
\pm m +1)(s \mp m -1)(s \pm m +2)$ and $\omega_{m,m\pm 2} = (E_m - E_{m\pm
2})/\hbar$. In Eq.~(\ref{eq:P}), $s_{-2}$ is equal $s_{-2}^m$, for $m=s$.

\begin{figure}
\fbox{\begin{minipage}[t]{0.5\textwidth}
  \centering
  \includegraphics[scale=0.70]{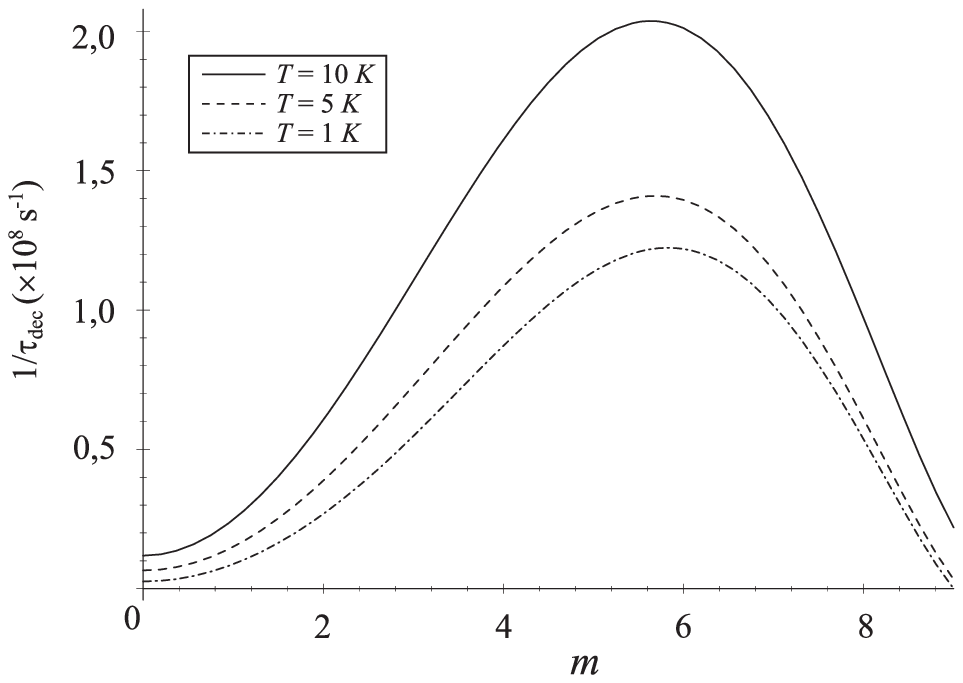}\\
  (a)
\end{minipage}}
\fbox{\begin{minipage}[t]{0.5\textwidth}
  \centering
  \includegraphics[scale=0.70]{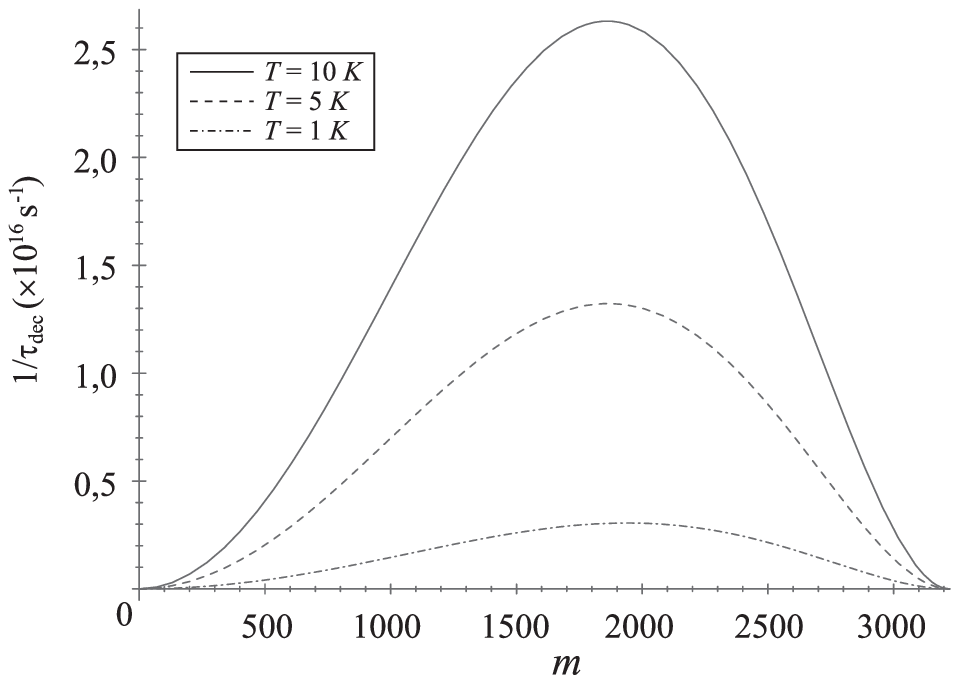}\\
  (b)
\end{minipage}}
\caption{\label{fig1} Inverse of the decoherence time as function of the initial state for the Mn$_{12}$ (a) and a typical superparamagnetic particle (b) at different temperatures.}
\end{figure}

Let us begin by analyzing the tunneling rate. The environmental effects are
contained in the brackets of Eq.~(\ref{eq:P}) and are, for typical values of the
constants (see Table~\ref{tab1}), much smaller than 1. We can therefore
conceive that the terms in brackets correspond to the first terms in the
expansion of the exponential $\e^{-\gamma(\beta)t}$, with
\[ \gamma(\beta) = \frac{18\Bint^2s_{-2}}{\hbar Mc^2\omega_D^3}
\frac{(\omega_{s,s-2})^3}{\e^{\hbar\omega_{s,s-2}/k_BT}-1}. \]
This result is thus very analogous to the tunneling rate obtained in the context of the spin-boson model, i.e., the ``free'' tunneling frequency is damped. Our model hypothesis, which we believe valid for magnetic particles, indicate that the tunneling rate remains essentially unaffected by the presence of the other degrees of freedom. This means, in particular, that the unitary quantum mechanical tunneling process will be hardly affected by this interaction. However, a careful look at the decoherence time (see Fig.~\ref{fig1}) shows that very quickly ($\tau_{\text{dec}} \sim 10^{-16}$s ($\sim 10^{-8}$s) for superparamagnetic (Mn$_{12}$) particles), many new channels will be open, the density matrix will be ``contaminated'' by other available states and lose purity. Physically this means that, although the particular coherences necessary for the macroscopic quantum coherent tunneling of the ground state remains essentially unaffected by the dissipation, states, candidate to tunnel the barrier, will not do it in a coherent way since others mechanisms for magnetization change will come into play very quickly, thus giving support to the idea that the experimentally observed magnetization change must be phonon assisted.

\begin{ack}
This is work was supported by Brazilian research agencies CNPq, FAPESP and FAPEMIG.
\end{ack}

\newpage

\begin{table}
\begin{center}
\caption{\label{tab1} Typical parameter values of the Mn$_{12}$ molecule and a superparamagnetic particle}
\begin{tabular}{lcc}
\hline\hline
Parameters & Mn$_{12}$ molecule$^{(1)}$ & Superparamagnetic particle$^{(2)}$ \\
\hline
$A$ & $7.5 \times 10^{-24}$J  & $2.57 \times 10^{-27}$ J \\
$B$ & $1.7 \times 10^{-26}$J  & $\approx 10^{-3}A$ \\
$B_{\text{int}}$ & $\approx A$ & $\approx 13$ cm$^{-1} = 4,1 \times 10^{-23}$J \\
$s$ & 10 & 3222 \\
density & $1.83 \times 10^{3}$ kg/m$^{3}$ & $5.0 \times 10^{3}$ kg/m$^{3}$ \\
sound's velocity & $2.0\times 10^{3}$ m/s & $3.0\times 10^{3}$ m/s \\
\hline\hline
\end{tabular}

Sources: ${}^{(1)}$\,Leuenberger and Loss \cite{leuenb3}; ${}^{(2)}$\,Pfannes et al.\cite{pfannes2}
\end{center}
\end{table}

\end{document}